\documentclass[preprint,11pt]{elsarticle}
\usepackage[hmargin=2.5cm,vmargin=2.5cm]{geometry}
\usepackage{array}

\usepackage{amssymb}
\usepackage{amsthm}
\usepackage{amsmath}
\usepackage{bm}
\usepackage{hyperref}
\usepackage{tcolorbox}
\hypersetup{
    colorlinks=true,
    linkcolor=blue,
    filecolor=magenta,      
    urlcolor=cyan,
    pdftitle={Overleaf Example},
    pdfpagemode=FullScreen,
    }
\usepackage{flafter}

\journal{Journal of Mechanics and Physics of Solids}

\begin{document}

\begin{frontmatter}



\title{A mechanically-derived contact model for adhesive elastic-perfectly plastic particles. Part II: Contact under high compaction---adding a bulk elastic response}


\author[inst1]{William Zunker}

\affiliation[inst1]{organization={Massachusetts Institute of Technology},
            addressline={77 Massachusetts Ave}, 
            city={Cambridge},
            postcode={02319}, 
            state={MA},
            country={USA}}

\author[inst1]{Ken Kamrin\corref{cor1}}
            \ead{kkamrin@mit.edu}
            \cortext[cor1]{Corresponding author.}

\begin{abstract}
In Part I of this two part series~\cite{zunker2023partI}, we presented a multi-neighbor dependent  contact model for adhesive elastic-plastic particles built upon the method of dimensionality reduction that is valid for the elastic and fully-plastic contact regimes. In this Part II, we complete the contact model by proposing a treatment for the bulk elastic contact regime which is characterized by a rapid stiffening in the force-displacement curve as interstitial pore spaces vanish. A simple formulation is presented for an additional bulk elastic force. A novel criterion for triggering this force (i.e. detecting the bulk elastic regime) related to the remaining free surface area of the particle is also given. This bulk elastic force is then superimposed with the force response given in Part I to achieve a contact model capable of capturing a variety of complex loadings. In this way, the methodology for treating the bulk elastic regime presented here stands independent of Part I and could be appended to any contact model. Direct comparison is made to finite element simulations showcasing bulk elastic responses, revealing the accurate predictive capabilities of the contact model. Notably, the contact model is also demonstrated to detect and evolve contacts caused purely by outward displacement of the free surface with good precision.     
\end{abstract}



\begin{keyword}
 B. contact mechanics \sep A. powder compaction \sep B. elastic–plastic material \sep A. adhesion  \sep high confinement bulk elastic response
\PACS 0000 \sep 1111
\MSC 0000 \sep 1111
\end{keyword}

\end{frontmatter}


\section{Introduction} \label{Introduction}

In Part I, the industrial relevance of powder compaction~\cite{samal2015powder,sigmund2000novel,meier2019modeling,ccelik2016pharmaceutical} and need for accurate numerical techniques was discussed. Emphasis was placed on the popular discrete element method (DEM)~\cite{cundall1979discrete}, as a choice method for many practitioners. The importance of contact models, describing the forces that arise between interacting particles, to the DEM was noted. Focus was directed towards normal contact models describing interacting spherical elastic-plastic particles and a review of them from the past few decades~\cite{chang1987elastic,storaakers1997similarity,mesarovic2000adhesive,zhao2000asperity,jackson2003finite,jackson2005finite,etsion2005unloading,harthong2009modeling,zait2010unloading,brake2012analytical,gonzalez2012nonlocal,agarwal2018contact,olsson2013force,frenning2013towards,frenning2015towards,brodu2015multiple,rathbone2015accurate,garner2018study,gonzalez2019generalized,edmans2020numerical,edmans2020unloading,giannis2021stress,giannis2021modeling,zhang2022research,xu2023new} revealed this to be an open and important problem. A generic force displacement curve for a contact between elastic-plastic particles in a packing was described. This led to the identification of three key regimes commonly cited in literature: elastic~\cite{hertz1882beruhrung}, fully-plastic~\cite{johnson1987contact}, and bulk elastic~\cite{tsigginos2015force}. The bulk elastic regime was discussed to be caused by a dramatic decrease of the pore space as the material tends towards a continuous medium. Correspondingly, this was shown to cause a rapid stiffening of the force displacement curve.

Of the reviewed contact models only four included treatment of the bulk elastic regime. The first was Harthong et al. in 2009~\cite{harthong2009modeling}. Although empirical in nature and lacking the ability to unload or model adhesion, this work was innovative in its usage of the Vorono{\"i} cell~\cite{gellatly1982calculation,gellatly1982characterisation} to predict the onset of the bulk elastic response. The major downside of this technique is that the Vorono{\"i} calculation adds significant computational expense especially for large simulations. Additionally, they only considered nearly rigid-plastic spheres, meaning the elastic deformation was negligible.  However in the case of non-negligible elastic deformations the volume of the sphere changes, which needs to be accounted for. 

In 2013 and 2015 Frenning published a series of work~\cite{frenning2013towards,frenning2015towards} outlining a contact model with a bulk elastic response. The developed model also lacks unloading and adhesive capabilities, but does present a geometrically and mechanically sound contact model with only one empirical parameter that can capture evolution of area, pressure, and force at each contact. It also avoids the usage of a Vorono{\"i} cell to trigger the bulk elastic response and instead relies on a simpler geometric criterion. The primary limitation of the model is that it is developed to only handle one specific loading case: triaxial compaction. 

Three years later Garner et al. in 2018 put forth a contact model equipped with a bulk elastic response as well as an ability to model unloading and adhesion.  Despite the range of features, the model is entirely empirical relying on numerous fitted functions to capture the force displacement curve. This means that calibration is needed each time the model is used for a new material. The contact model is also fitted to a specific coordination number meaning all contacts, no matter the loading configuration, produce the same force-displacement curves. Correspondingly, the bulk elastic response is always triggered at the same time; at the grain-scale level this could produce poor results since coordination number may vary for each particle. 

More recently in 2021 Giannis et al.~\cite{giannis2021modeling} published a contact model including a bulk response, unloading, and adhesion. Like Garner et al. this contact model is empirical relying on fitted spring constants leading to the same calibration issues when considering different materials. One interesting addition to the contact model is their treatment of multi-neighbor dependent  effects which relies on analysis of the trace of the stress tensor for each particle using the method described in~\cite{giannis2021stress}. Nevertheless, the actual evolution of the bulk response is given largely through an empirically fitted function and is triggered by criteria on the force.      

 Review of the previous treatments of the bulk elastic response has revealed that the response in this regime has historically been treated through some empirical fit or, in the case of Frenning's work, for only one loading type. Additionally, understanding of when to trigger the bulk response is highly variant---ranging from the use of a Vorono{\"i} cell to simple inequalities on the force. This leaves space for improvement, with an ideal bulk response being triggered by a mechanism related to the reduction of the pore space and evolution of the force grounded in solid mechanics. 

 In what follows we will outline a treatment that addresses both improvements. To begin, the method of dimensionality reduction (MDR) contact model from Part I will be succinctly summarized in Section~\ref{Method of dimensionality reduction (MDR) contact model summary}. Finite element simulations on single elastic-plastic particles under various loading conditions (triaxial compaction, die compaction, and uniaxial compression) will be carried out in Section~\ref{Finite element simulations showcasing a bulk elastic response}. This will exhibit the rich contact behavior displayed for various loading types and motivate the need for a flexible bulk response as developed in Section~\ref{Adding a bulk elastic response to the MDR contact model}. Comparison between the MDR contact model with a bulk response and finite element simulations will be shown in Section~\ref{Verification of the MDR contact model with a bulk elastic response} highlighting the robust capabilities of the full contact model.  

\section{Method of dimensionality reduction (MDR) contact model summary} \label{Method of dimensionality reduction (MDR) contact model summary}

The non-adhesive MDR contact model is summarized below. It is based on the MDR methodology developed by Popov and He{\ss}~\cite{popov2013methode,popov2015method}. For reference all the important model parameters are tabulated in Table~\ref{nonadhesive MDR parameters}. The force is given as a function of the apparent overlap $\delta$, maximum experienced apparent overlap $\delta_\textrm{max}$, and apparent radius $R$

    \begin{equation} 
        	 F(\delta,\delta_\textrm{max},R) = \frac{E^*_c AB}{4}\left[ \arccos\left( {1 - \frac{2\delta^e_{\textrm{1D}}}{A}}\right) - \left( 1 - \frac{2\delta^e_{\textrm{1D}}}{A} \right) \sqrt{\frac{4\delta^e_{\textrm{1D}}}{A} - \frac{4(\delta^e_{\textrm{1D}})^2}{A^2}}\right].
    \end{equation}

\noindent In the elastic regime, before the onset of plasticity, $\delta^e_\textrm{1D} = \delta$. The composite plane strain modulus of the two contacting bodies $B_1$ and $B_2$, each with their respective Young's modulus $E_i$ and Poisson's ratio $\nu_i$ is given as

  \begin{equation} \label{Ecomposite}
	E^*_c = \left( \frac{1-\nu_{1}^2}{E_{1}} + \frac{1-\nu_{2}^2}{E_{2}} \right)^{-1}.
\end{equation}

\noindent The apparent overlap and maximum experienced apparent overlap are `displacements' measured from a spherical surface of radius $R$, and are related to the displacement $\delta_o$ and maximum experienced displacement $\delta_{o,\textrm{max}}$, which are displacements measured with respect to the initial radius $R_o$, through the following relations

\begin{equation} \label{current deltas}
    \delta = \delta_o + R - R_o, \qquad \delta_\textrm{max} = \delta_{o,\textrm{max}} + R - R_o. 
\end{equation}

\noindent The parameters $A$ and $B$ define an elliptical indenter in the transformed space\footnote{The elliptical indenter in the transformed space corresponds exactly to a known pressure distribution of the contact in 3D space.} and are given as

    \begin{equation}
        \begin{split}
            & A = 4R, \qquad \qquad \; \; B = 2R, \qquad \quad \; \textrm{elastic} \\
            & A = \frac{4p_Y}{E^*_c}a_\textrm{max}, \qquad B = 2a_\textrm{max}, \qquad \textrm{fully-plastic}.
        \end{split}
    \end{equation}

\noindent The average pressure within a contact in the fully-plastic regime $p_Y$ is defined by the average pressure for the limiting case of a rigid-plastic sphere

    \begin{equation} \label{pYfit}
        	p_Y = Y\left( 1.75\exp{(-4.4\delta_\textrm{max}/R)+1} \right).
    \end{equation}

\noindent The transition between the elastic and fully-plastic regimes is taken to be sharp and is based on the average pressure at the contact. Namely, once the average pressure described by Hertz's contact law

    \begin{equation} \label{pbarhertz}
        	 \bar{p}_H = \frac{4E^*_c}{3 \pi \sqrt{R}}\sqrt{\delta},
    \end{equation}

\noindent meets that of the hardening curve (\ref{pYfit}) the formulation switches from elastic to fully-plastic. The transformed 1D elastic displacement is given by 

    \begin{equation} 
        	 \delta^e_\textrm{1D} = \frac{\delta - \delta_\textrm{max} + \delta^e_\textrm{1D,max} + \delta_R}{1 + \delta_R/\delta^e_\textrm{1D,max}},
    \end{equation}

\noindent where

    \begin{equation} 
        	 \delta^e_\textrm{1D,max} = A/2,
    \end{equation}

\noindent the displacement correction is given as

    \begin{equation} 
        	 \delta_R = \frac{F(\delta_\textrm{max},\delta_\textrm{max},R)}{\pi a_\textrm{max}^2} \left[  \frac{2a^2_\textrm{max} (\nu - 1) - z_R(2\nu - 1)( \sqrt{a^2_\textrm{max} + z^2_R} - z_R)} {2G\sqrt{a_\textrm{max}^2 + z_R^2}} \right],  
    \end{equation}

\noindent and the distance from the free surface center of the sphere is

    \begin{equation}
        	 z_R = R - (\delta_\textrm{max} - \delta^e_\textrm{1D,max}). 
    \end{equation}

\noindent The maximum experienced contact radius is  

        \begin{equation}
        	 a_\textrm{max} = \sqrt{(2\delta_\textrm{max} R - \delta^2_\textrm{max}) + c_A/\pi},
    \end{equation}

\noindent where $c_A$ is found to enforce continuity of the contact radius between the elastic and plastic regimes

    \begin{equation}
        	 c_A = \pi(\delta_Y^2 - \delta_Y R).
    \end{equation}

\noindent The yield displacement $\delta_Y$ is found by equating $\bar{p}_H$ and $p_Y$ 

    \begin{equation} 
        	 \frac{4E^*_c}{3 \pi \sqrt{R}}\sqrt{\delta_Y} = Y\left( 1.75\exp{(-1.1\delta_Y/R)+1} \right).
    \end{equation} 
    
\noindent Finally, the radius is allowed to grow to respect the incompressible nature of the plastic deformation, giving the incremental form

\begin{equation}
    \Delta R = \textrm{max} \left[ \frac{\Delta V^e - \sum_{i=1}^{N} \pi\Delta \delta_{o,i}(2\delta_{o,i} R_o - \delta^2_{o,i} + R^2 - R^2_o) }{2 \pi R \sum_{i=1}^{N}(\delta_{o,i} + R - R_o) - 4 \pi R^2}, \; 0 \right].
 \end{equation}

\noindent The elastic volume change $\Delta V^e$ (taken as positive for a compressive increment) is determined from the volume change between steps, where the total volume $V$ is calculated as

        \begin{equation} \label{elastic volume}
    	V = V_o \left( 1 + \frac{1}{3\kappa V}\textrm{tr} \left( \sum_{c \; \in \; V} \bm{f}^c \otimes \bm{b}^c \right) \right).
        \end{equation}

Adhesion can be appended to this contact law using the methodology outlined in Part I.

\begin{table} [htp] 
\begin{center}
        \renewcommand{\arraystretch}{1.25}
	\begin{tabular}{ ||c|l|| } 
		\hline
		$A$ & height of elliptical indenter \\ \hline
            $a$ & contact radius \\ \hline
            $a_\textrm{max}$ & maximum experienced contact radius \\ \hline
            $B$ & width of elliptical indenter \\ \hline
            $\bm{b}$ & branch vector \\ \hline
            $c_A$ & contact area intercept \\ \hline
            $\delta_o$ & displacement \\ \hline
            $\delta_{o,\textrm{max}}$ & maximum experienced displacement \\ \hline
            $\delta$ & apparent overlap \\ \hline
            $\delta_\textrm{max}$ & maximum experienced apparent overlap \\ \hline
            $\delta^e_\textrm{1D}$ & transformed elastic displacement \\ \hline
            $\delta^e_\textrm{1D,max}$ & maximum transformed elastic displacement \\ \hline
            $\delta_R$ & displacement correction \\ \hline
            $\delta_Y$ & yield displacement \\ \hline
            $E$ & Young's modulus \\ \hline
            $E^*_c$ & composite plane strain modulus \\ \hline
            $F_\textrm{MDR}$ & force from MDR \\ \hline
            $\bm{f}$ & force vector \\ \hline
            $G$ & shear modulus \\ \hline
            $\kappa$ & bulk modulus \\ \hline
            $\nu$ & Poisson ratio \\ \hline
            $\bar{p}_H$  & average pressure for Hertz contact \\ \hline
            $p_Y$  & average pressure along hardening curve \\ \hline
            $R_o$ & initial radius \\ \hline
            $R$ & apparent radius \\ \hline
            $\Delta R$ & change in apparent radius \\ \hline
            $V_o$ & initial volume \\ \hline
            $V$ & current volume \\ \hline
            $\Delta V^e$ & change in elastic volume \\ \hline
            $Y$ & yield stress \\ \hline
            $z_R$ & depth of particle center \\ 
		\hline
	\end{tabular}
\end{center}
\caption{Important parameters for evaluation of non-adhesive MDR contact model.}
\label{nonadhesive MDR parameters}
\end{table}

\section{Finite element simulations showcasing a bulk elastic response} \label{Finite element simulations showcasing a bulk elastic response}

\subsection{Simulation setups}

\begin{figure*} [!htb]
 	\raggedright
 	\includegraphics[width=\textwidth, trim = 0cm  10cm 0cm 0cm]{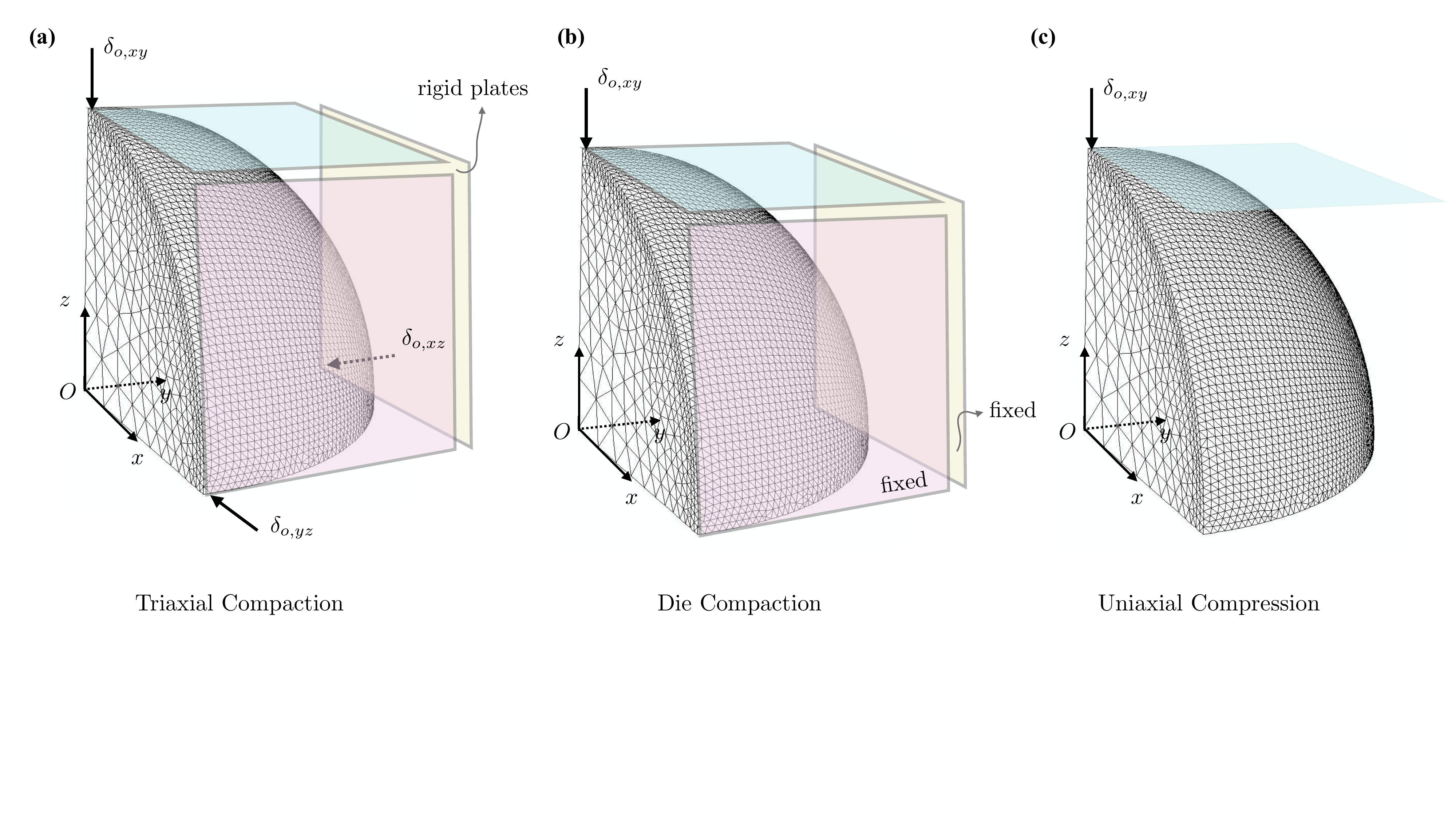}
 	\caption{Finite element simulation set ups for three different loading configurations: (a) triaxial compaction, (b) die compaction, and (c) uniaxial compression.}
 	\label{fem_bulk_setup}
 \end{figure*}

Loadings that give rise to a bulk elastic response commonly occur in applications such as powder compaction. To probe the behavior of individual contacts for these types of loadings and to provide guidance for appending a bulk response into our contact model, we perform three types of finite element simulations with the software Abaqus involving the same three loadings as investigated in~\cite{harthong2009modeling}: triaxial compaction, die compaction, and uniaxial compression as shown in Fig.~\ref{fem_bulk_setup}(a)-(c). We consider contact between a single spherical particle of initial radius $R_o$ and rigid flats. As in Part I, the material is isotropic, elastic perfectly-plastic obeying a von Mises yield condition, and homogeneous with a Young's modulus $E$, Poisson ratio $\nu$, and yield stress $Y$. No variation in $E/Y$ or $\nu$ is considered and they are held fixed at values of $20$ and $0.3$, respectively. The value of $E/Y=20$ is selected to produce a non-negligible elastic regime. 

Fully 3D simulations are considered using a one-eighth sphere geometry. Symmetry boundary conditions restricting appropriate displacement degrees of freedom along the $xy$, $xz$, and $yz$ faces are enforced for equivalence with a full sphere geometry. A mesh refinement study led to a total of 48,167 C3D10 (10-node quadratic tetrahedron) elements. 

In the case of triaxial compaction Fig.~\ref{fem_bulk_setup}(a), three rigid flats parallel to the coordinate planes $xy$, $xz$, and $yz$ are placed in contact with the eighth sphere at its three exterior points that lie along the coordinate axes. To drive the simulation displacements $\{ \delta_{o,xy}, \delta_{o,xz}, \delta_{o,yz} \}$ directed along the coordinate axes are applied to the respective rigid planes displacing them towards the sphere center. All displacements are equal to $\delta_o$ resulting in a uniform compression of the sphere along all three axes. 

The case of die compaction, Fig.~\ref{fem_bulk_setup}(b), has the same initial set up as triaxial compaction. The difference lies in how the simulation evolves, in this case the $xz$ and $yz$ rigid flats are held fixed, whereas the $xy$ flat is driven as before. The contact that forms with the $xy$ flat is denoted as a primary contact. Global deformations induced by this primary contact result in radial expansion of the sphere giving rise to secondary contacts at the $xz$ and $yz$ planes. 

The last simulation is uniaxial compression Fig.~\ref{fem_bulk_setup}(c), here, the $xz$ and $yz$ rigid flats are removed entirely leaving only the $xy$ flat. This simulation is exactly identical to those performed in Part I, providing a baseline comparison to what has been previously presented.

\subsection{Simulation results}

In the FEM results four total curves are present for the force, area, and average contact pressure evolution since the primary and secondary contact behavior for die compaction are independently tracked. During the die compaction simulation the rigid flats at the secondary contacts are held fixed; the displacement used for plotting the secondary contacts is therefore $\delta_{o,xy}$. For the volume calculation only three curves are present for each type of loading considered.

\subsubsection{Force-displacement}

\begin{figure*} [!htb]
 	\centering
 	\includegraphics[width=\textwidth, trim = 0cm  1.5cm 9cm 2cm]{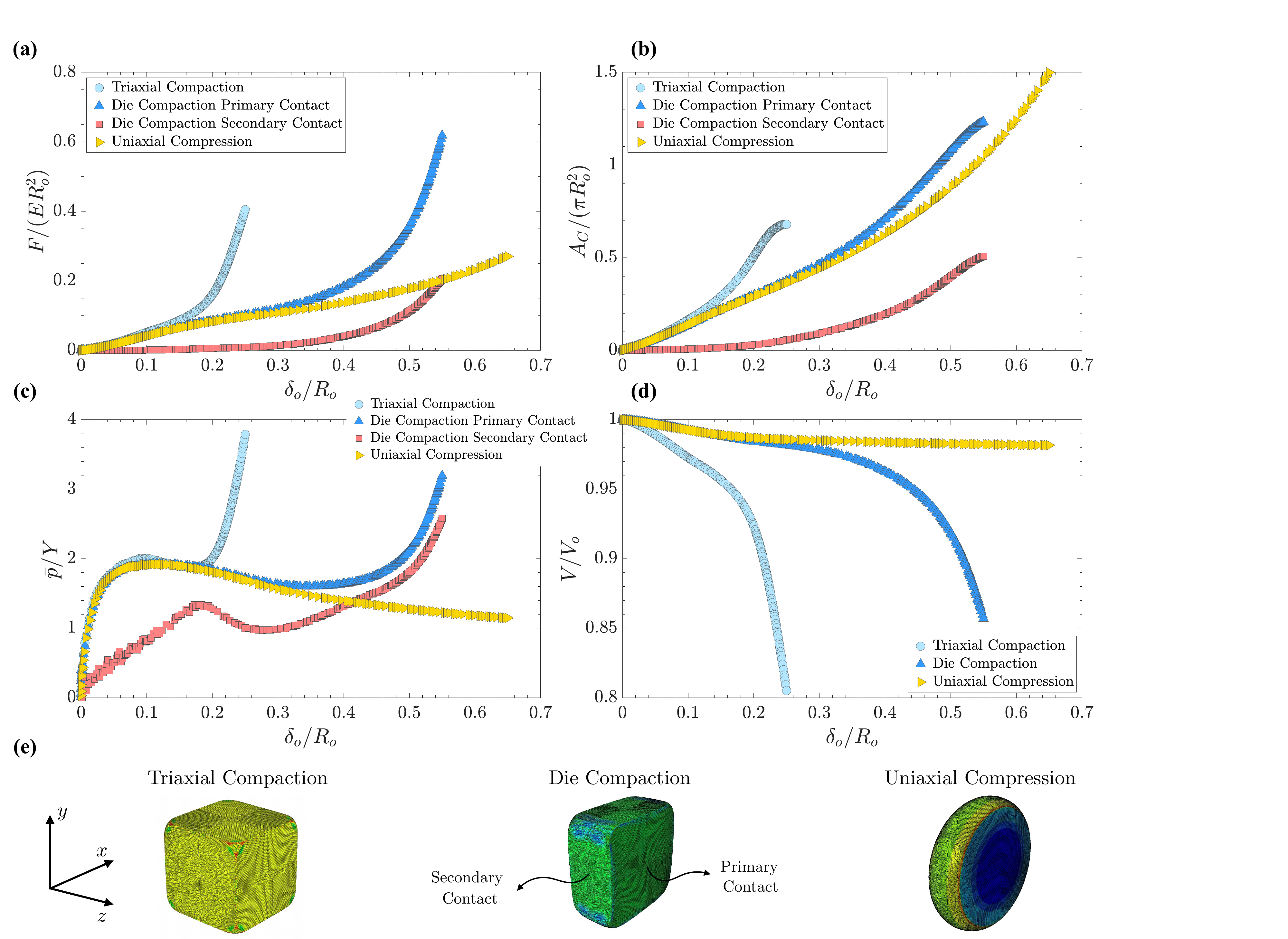}
 	\caption{Finite element method simulation results showcasing bulk elasticity responses. (a) Nondimensionalized curves for force-displacement, (b) contact area-displacement, (c) average contact pressure-displacement, and (d) particle volume-displacement. (e) Final deformed configurations of the sphere for each case---colors can be ignored and are there to help visualization of the shapes.}
 	\label{fem_bulk_results}
 \end{figure*}

The normalized force-displacement curves for the three different loadings are shown in Fig.~\ref{fem_bulk_results}(a).  Under small deformations $\delta_o/R_o < 0.1$ we observe that the evolution of triaxial compaction, die compaction primary contact, and uniaxial compression are essentially identical following Hertz's contact law until plastic deformation begins. The die compaction secondary contact has a negligible force response at this stage. This provides evidence that all contacts can initially be seen to evolve independently under small deformation\footnote{This fact can be interpreted through the lens of Saint Venant's principle---provided that the local deformations and stresses caused by contacts are small, they decay rapidly enough to not significantly influence other contacts.  Rather the other contacts only act to constrain the body from translating keeping it in static equilibrium, but the exact manner of constraint is not important.}.    

In considering larger displacements $\delta_o/R_o > 0.1$ we first note that the case of uniaxial compression does not show a bulk response due to lack of confinement and thus provides a baseline for comparison. \textit{Although the uniaxial compression case does not show a bulk response there is multi-neighbor dependent  behavior caused by interaction between the two contacts.} Moving beyond $\delta_o/R_o > 0.1$ triaxial compaction is the first to deviate from uniaxial compression. This is caused by an early triggering of significant multi-neighbor dependent  effects due to the higher number of contacts on the particle. By $\delta_o/R_o = 0.2$ the force-displacement curve increases greatly in slope, marking the full transition to a bulk response. At this point, the deformed sphere begins to look like the cube shown in Fig.~\ref{fem_bulk_results}(e). We observe that the body has essentially reduced to the classically drawn infinitesimal continuum element, with roughly uniform stresses applied to each face. With all stresses being equal on each face and lacking a shear component (i.e. a spherical stress state) we, unsurprisingly, see a bulk elastic response related back to the bulk modulus $\kappa$. A similar behavior is found for the case of die compaction, the key difference is in the delayed onset of this bulk response, caused by the static rather than active confinement at the secondary contacts. The final shape of the particle also reflects the specifics of the loading, ending as a rectangular prism rather than a cube. The force at the die compaction secondary contact remains comparatively small to the primary contact until the transition to the bulk response begins, thereafter following the same trend as the primary contact.


\subsubsection{Contact area-displacement}

The trends seen for the force-displacement curves are well reflected in the contact area-displacement curves shown in Fig.~\ref{fem_bulk_results}(b). Initially, we see that the cases of triaxial compaction and die compaction primary contact match with uniaxial compression, which follows Hertz's contact law initially before plastic deformation. Deviation from uniaxial compression begins at the same points for both contacts. As with the force, the die compaction secondary contact area remains comparatively small to the primary contact until the transition to the bulk response. One unique feature about the contact area evolution is the stagnation of growth in area that happens during the bulk response, observable in both the case of triaxial and die compaction. This stagnation is consistent with the idea that under high confinement there is minimal pore space for material to flow into, thus area does not accumulate as freely as before. 

\subsubsection{Average contact pressure-displacement}

As before, triaxial compaction and the die compaction primary contact follow the uniaxial compression case almost exactly until the transition to the bulk response initiates as shown in Fig.~\ref{fem_bulk_results}(c). At larger deformations the average pressure for uniaxial compression trends towards the value of the yield stress $Y$. This is caused by the deformed geometry seen in Fig.~\ref{fem_bulk_results}(e) beginning to resemble a cylinder more closely than a sphere, a phenomenon first noted in~\cite{quicksall2004elasto,jackson2005finite}. The cases of triaxial compression and die compaction on the other hand have rising average pressures once the bulk response initiates. The die compaction secondary contact behavior can be mostly ignored from $0 \leq \delta_o/R_o \leq 0.25$, during this period the force and contact area are very small making them prone to numerical errors\footnote{The primary error arises from the discretization of the mesh being too small to make out the small variations in the area. This causes the area to remain constant until another element makes contact with the flat leading to a non-smooth contact area evolution.}. Beyond this region the force and area quickly become of the same order as the primary contact and the assigned mesh becomes well adapted to resolving the evolution of the average contact pressure.

\subsubsection{Volume-displacement}

The particle volume-displacement relation is shown in Fig.~\ref{fem_bulk_results}(d), where the volume $V$ is normalized by the initial volume $V_o = 4/3\pi R^3_o$. All volume loss is elastic due to the incompressibility of the plastic deformations. Triaxial compaction is observed to reduce in volume the most rapidly---a result that makes physical sense considering the high confinement. Die compaction and uniaxial compression are observed to have similar reductions in volume until approximately $\delta_o/R_o > 0.2$, after which, die compaction begins to accumulate more volume loss. The most rapid volume loss occurs in both triaxial compaction and die compaction once their respective bulk responses turn on. This is due to the large spherical stress in the bulk elastic regime that leads to significant elastic volumetric strains.   

\section{Adding a bulk elastic response to the MDR contact model} \label{Adding a bulk elastic response to the MDR contact model}

Adding a bulk elastic response to the MDR contact model summarized in Section~\ref{Method of dimensionality reduction (MDR) contact model summary} is straightforward, but requires additional kinematic bookkeeping and development of a criterion to trigger the bulk response. A key observation is that elastic-plastic solutions can be superimposed with a bulk elastic response. Superposition holds because a bulk elastic response can be approximated as only affecting the spherical stress state, leaving the deviatoric\footnote{A von Mises yield criterion is used for all finite element simulations.} part that controls yielding unchanged.  

\subsection{Criterion for onset of bulk elastic response}

The FEM study from Section~\ref{Finite element simulations showcasing a bulk elastic response} shows that the bulk elastic response initiates at different displacements for different loading configurations. A physically motivated criterion related to the reduction of pore space to predict the onset of the bulk response is needed. In place of the previously used Vorono{\"i} cell calculation~\cite{harthong2009modeling}, we introduce a new criterion that simply compares the free surface area not part of any contact $A_\textrm{free}$ to the total surface area $A_\textrm{tot}$. If the ratio drops below a specified value $\psi_b$ then the bulk response turns on; that is, when

        \begin{equation} \label{AfreeAtotRatio}
    	\frac{A_\textrm{free}}{A_\textrm{tot}} < \psi_b. 
        \end{equation}

\noindent No precise value of $\psi_b$ exists, however $\psi_b$ around $0.08$ gives reasonable results. A nice feature of this criterion is that it is based on the already known surface areas.

\subsection{Kinematic and force decompositions}

Before discussing the additional kinematics needed for the bulk response it is useful to recall the two important types of kinematic measures already in place from Part I: 1) the displacements measured with respect to $R_o$ including the displacement $\delta_o$ and maximum experienced displacement $\delta_{o,\textrm{max}}$ and 2) the `displacements' measured with respect to $R$ including the apparent overlap $\delta$ and maximum experienced apparent overlap $\delta_\textrm{max}$. It is simple to convert between these kinematic measures, provided the apparent radius $R$ is known, as shown in (\ref{current deltas}).   

In developing the bulk elastic response we are guided by the idea of superposition of elastic states, viewing the boundary tractions as those of an accruing bulk uniform compaction, plus those attributable to the MDR contact model.  This naturally leads to a new\footnote{It is important to recall the already existing kinematic decomposition of the MDR portion of the displacement into an elastic and plastic portion $\delta_{o,\textrm{MDR}} = \delta^e_{o,\textrm{MDR}} + \delta^p_{o,\textrm{MDR}}$.} kinematic decomposition of the displacement $\delta_o$ into a MDR and bulk component
        \begin{equation} \label{delta_decompostion}
    	\delta_o = \delta_{o,\textrm{MDR}} + \delta_{o,\textrm{Bulk}}. 
        \end{equation}

\noindent The MDR portion of the displacement is fed-in directly to the model summarized in Section~\ref{Method of dimensionality reduction (MDR) contact model summary} in place of $\delta_o$ (i.e. $\delta_o = \delta_{o,\textrm{MDR}}$). The resulting force given from that process, $F_\textrm{MDR}$, will be added to the force from the bulk response, $F_\textrm{Bulk}$, which depends on $\delta_{o,\textrm{Bulk}}$ to return the total force at a given contact
        \begin{equation} \label{F_total}
    	\boxed{F = F_\textrm{MDR} + F_\textrm{Bulk}}. 
        \end{equation}

\noindent In the following section we develop a general formulation for the bulk force.

\subsection{General bulk force formulation}

The increments in displacement $d\delta_o$ at a contact are partitioned depending on the ratio of $A_\textrm{free}/A_\textrm{tot}$  
\begin{equation} \label{ddelta_oMDR and ddelta_oBulk}
    \begin{split}
        & d  \delta_{o,\textrm{MDR}} = d \delta_o, \qquad \qquad \; \; d \delta_{o,\textrm{Bulk}} = 0 \qquad \quad \textrm{if} \; A_\textrm{free}/A_\textrm{tot} \geq \psi_b, \\
        & d  \delta_{o,\textrm{MDR}} = d \delta_o - d \bar{\delta}_o, \qquad d \delta_{o,\textrm{Bulk}} = d\bar{\delta}_o \qquad \,\textrm{otherwise}.
    \end{split}
\end{equation}

\noindent The upper case of (\ref{ddelta_oMDR and ddelta_oBulk}) corresponds to when the criterion for the bulk elastic response is \textit{not} met. The lower case pertains to instances when $A_\textrm{free}/A_\textrm{tot} < \psi_b$, associated with an active bulk elastic response. Albeit not obvious at first, during an active bulk response, an increment of displacement is split by apportioning the mean surface displacement across all contacts $d \bar{\delta}_o$ to $\delta_{o,\textrm{Bulk}}$, the residual or deviator $d \delta_o - d\bar{\delta}_o$ is then assigned to $\delta_{o,\textrm{MDR}}$ as shown in (\ref{ddelta_oMDR and ddelta_oBulk}).




To justify this displacement decomposition during an active bulk response, we recall that for an elastic-plastic particle with incompressible plastic flow, the volume-averaged volumetric strain\footnote{The traceless property of the volume-averaged infinitesimal rotation tensor $\bar{\bm{\omega}}$ is used here: $\textrm{tr}(\overline{\nabla \bm{u}}) = \textrm{tr}(\bar{\bm{\epsilon}} + \bar{\bm{\omega}}) = \textrm{tr}(\bar{\bm{\epsilon}})$, where $\bar{\bm{\epsilon}}$ is the volume-averaged infinitesimal strain tensor.} is proportional to the volume-averaged mean normal stress
        \begin{equation} \label{nabla u bar mean stress bar}
    	\textrm{tr}(\overline{\nabla \bm{u}}) = \frac{\textrm{tr}(\bm{\bar{\sigma}})}{3\kappa}
        \end{equation}

\noindent where $\kappa$ is the bulk modulus, $\bm{\bar{\sigma}}$ the volume-averaged Cauchy stress tensor, and $\overline{\nabla \bm{u}}$ the volume-averaged gradient of the vector-valued displacement field $\bm{u}$, which can be expressed as

        \begin{equation} \label{nabla u surface integral}
        \overline{\nabla \bm{u}} = \frac{1}{V} \int_S \bm{n} \otimes \bm{u} \,dS,	
        \end{equation}

\noindent where $\bm{n}$ is the surface normal.

We now consider a particle that has entered the bulk elastic regime, with N contacts each with an area $A_{C,i}$ and normal $\bm{n}_i$ that each experience a vector-valued incremental displacement $d\bm{\delta}_{o,i}$. Under the assumption that the free surface area of the particle is now negligible, $A_\textrm{free}/A_\textrm{tot}\ll 1$, an increment of (\ref{nabla u surface integral}) approximates to the discrete form

        \begin{equation} \label{nabla u surface integral discrete}
       d (\overline{\nabla \bm{u}}) = \frac{1}{V} \int_S \bm{n} \otimes d\bm{u} \,dS \approx \frac{1}{V} \sum_{i=1}^{N} (A_{C,i} \bm{n}_i \otimes d\bm{\delta}_{o,i}).	
        \end{equation}

\noindent Inserting (\ref{nabla u surface integral discrete}) into (\ref{nabla u bar mean stress bar}) gives

        \begin{equation} \label{discrete nabla bar u mean stress bar}
        \textrm{tr} \left( \frac{1}{V} \sum_{i=1}^{N} (A_{C,i} \bm{n}_i \otimes d\bm{\delta}_{o,i}) \right) = \frac{\textrm{tr}(d\bm{\bar{\sigma}})}{3 \kappa}.	
        \end{equation}

\noindent Factoring the $1/V$ outside the trace operation and defining $A_\textrm{con}$ to be equal to the total surface area involved in the contacts as well as using the identity $\textrm{tr}(\bm{v} \otimes \bm{w}) = \bm{v} \cdot \bm{w}$ (\ref{discrete nabla bar u mean stress bar}) can be rewritten as

        \begin{equation} \label{Acon discrete nabla bar u mean stress bar}
        \frac{A_\textrm{con}}{V} \left( \frac{1}{A_\textrm{con}} \sum_{i=1}^{N} (A_{C,i} \bm{n}_i \cdot d\bm{\delta}_{o,i}) \right) = \frac{\textrm{tr}(d\bm{\bar{\sigma}})}{3 \kappa}.	
        \end{equation}

\noindent Inspection of the above expression reveals that it includes a term equal to the mean surface displacement increment across all contacts

        \begin{equation} \label{Delta delta bar}
         d\bar{\delta}_o =  \frac{1}{A_\textrm{con}} \sum_{i=1}^{N} (A_{C,i} \bm{n}_i \cdot d\bm{\delta}_{o,i}).	
        \end{equation}

\noindent This allows (\ref{discrete nabla bar u mean stress bar}) to be further rewritten as 
        \begin{equation} \label{mean stress mean displacement}
    	\frac{\textrm{tr}(d\bm{\bar{\sigma}})}{3} = \frac{A_\textrm{con}}{V}\kappa\,  d\bar{\delta}_o .
        \end{equation}

\noindent Relation (\ref{mean stress mean displacement}) shows that the mean surface displacement is directly connected to the bulk response, justifying the displacement decomposition in the lower case of (\ref{ddelta_oMDR and ddelta_oBulk}). 

Finally, we claim that increments in volumetric strain that occur during the bulk elastic regime are caused purely by increments in the bulk contact force $dF_\text{Bulk}$.  Hence, $dF_{\text{Bulk},i}$, at a given contact $i$, is simply given by multiplying the hydrostatic part of the volume-averaged stress increment with the contact area of the given contact 
\begin{equation}
    dF_{\text{Bulk},i}=\frac{\textrm{tr}(d\bm{\bar{\sigma}})}{3} A_{C,i} = \frac{A_\textrm{con}}{V} \, \kappa \, d\bar{\delta}_o  \, A_{C,i}.
\end{equation}
Integration then provides
\begin{equation} \label{Fbulk}
    F_{\text{Bulk},i}=\int dF_{\text{Bulk},i},\ \qquad \delta_{o,\text{Bulk},i}=\int d\bar{\delta}_o,
\end{equation}
recalling that the integrands here are only non-zero after entering the bulk elastic state, $A_\textrm{free}/A_\textrm{tot} < \psi_b$;  prior to this, all increments in the bulk displacement are zero based on the upper case of (\ref{ddelta_oMDR and ddelta_oBulk}). 

With (\ref{Fbulk}) established we have an expression for calculating the force at each contact due to the bulk response. To close the expression, ways to determine $A_{C,i}$, $A_\textrm{con}$, and $V$ need to be detailed. 

\subsubsection{Deformed volume in the bulk elastic regime}

The volume $V$ in (\ref{Fbulk}) is taken to be given by a purely geometric picture
    \begin{equation} \label{Vdeformed}
        	 V_\textrm{geo} = \frac{4}{3}\pi R^3 - \sum_{i=1}^{N} \left( \frac{\pi}{3} \delta^2_i[3R - \delta_i] \right).
    \end{equation}

\noindent This same expression was used successfully to derive the radius growth scheme introduced in Part I. The first term corresponds to the geometric volume of a sphere with the apparent radius, and the second term to the spherical caps defined by the intersections with the rigid flats. In this way, the deformed volume is calculated as that of a sphere with chopped off caps at each of the contacts.

\subsubsection{Contact area, total area, and free area in the bulk elastic regime}

 The contact area for each contact during the bulk response uses the same general geometric intersection form as in the fully-plastic regime for the MDR contact model during forward loading with $\delta_i$ in place of $\delta_\textrm{MDR,i}$
        \begin{equation} \label{A_{C,i}}
    	A_{C,i} = \pi (2 \delta_i R - \delta^2_i) + c_{A,i}.
        \end{equation}

\noindent The total area involved in contacts is simply the summation of each individual contact
        \begin{equation} \label{A_con}
    	A_\textrm{con} = \sum_{i=1}^{N} A_{C,i}.
        \end{equation}

\noindent With (\ref{A_con}) established the total particle area is given as
\begin{equation} \label{Atot}
    A_\textrm{tot} = 4\pi R^2 - 2\pi \sum_{i=1}^{N} (\delta_\textrm{max,MDR,i} + \delta_{o\textrm{,Bulk,i}})R + A_\textrm{con}.
\end{equation}

\noindent Expression (\ref{Atot}) can be interpreted as taking the full sphere surface area, subtracting off the areas of the spherical caps intersected by the rigid flats, and then adding back the total area of the contacts. The free area not involved in contact is then naturally defined as
\begin{equation} \label{Afree}
    A_\textrm{free} = A_\textrm{tot} - A_\textrm{con}.
\end{equation}

\subsubsection{Stopping radius growth in the bulk elastic regime} \label{Stopping radius growth in the bulk elastic regime}

In Part I, the radius growth scheme was introduced, this led to the following differential update form for the apparent radius 

\begin{equation} \label{DeltaR}
    \Delta R = \frac{\Delta V^e - \sum_{i=1}^{N} \pi\Delta \delta_{o,i}(2\delta_{o,i} R_o - \delta^2_{o,i} + R^2 - R^2_o) }{2 \pi R \sum_{i=1}^{N}(\delta_{o,i} + R - R_o) - 4 \pi R^2}.
 \end{equation}

\noindent In the development of (\ref{DeltaR}), it was remarked that under highly confined conditions erroneous radius growth may be predicted. This is linked back to the fundamental assumption used to derive $\Delta R$:

\begin{itemize}
    \item the deformed volume is well approximated by a sphere with chopped off spherical caps where the rigid flats intersect.
\end{itemize} 

\noindent Under high confinement this assumption is violated because the spherical cap volumes predicted from the intersection with the rigid flats begin to severely intersect one another---a second order geometric effect that is not accounted for in (\ref{DeltaR}). This leads to non-physical and incorrect predictions of $\Delta R$ including negative or anomalously large values. We recall that the bulk elastic regime is only triggered when $A_\textrm{free}/A_\textrm{tot} < \psi_b$, which means that for low values of $\psi_b$ (e.g. 0.08) the bulk elastic regime begins under reasonably high confinement. \textit{To prevent non-physical predictions of the radius growth from occurring the radius growth is halted at the beginning of the bulk elastic regime, but may resume if the bulk elastic regime is exited.}      

\section{Verification of the MDR contact model with a bulk elastic response} \label{Verification of the MDR contact model with a bulk elastic response} 

\subsection{Force-displacement comparison}

\begin{figure*} [!htb]
 	\centering
 	\includegraphics[width=\textwidth, trim = 1cm  2.5cm 13cm 1cm]{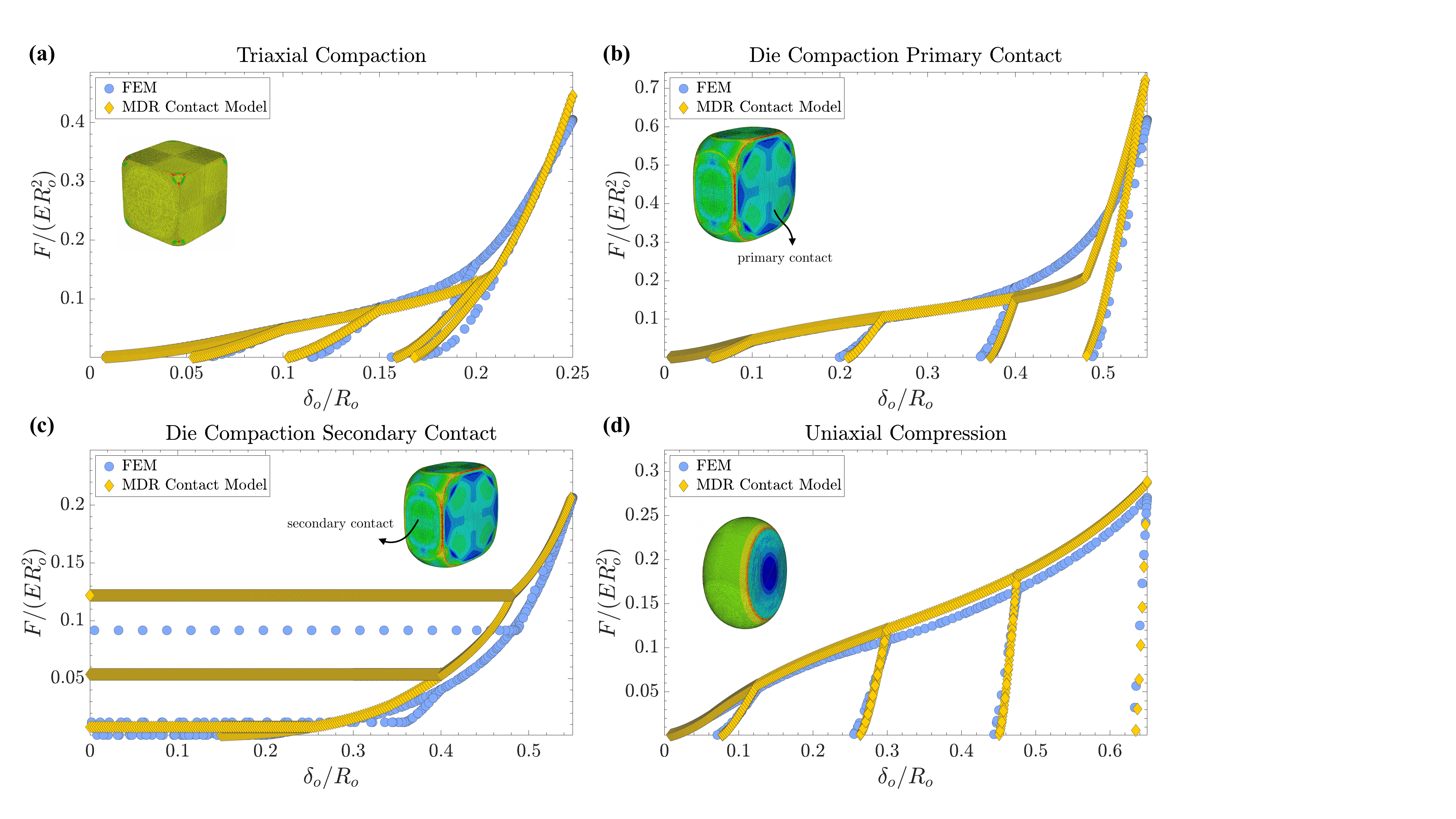}
 	\caption{Force-displacement curves for the four types of contacts studied: (a) triaxial compaction, (b) die compaction primary contact, (c) die compaction secondary contact, and (d) uniaxial compression.}
 	\label{bulk_fem_mdrcm_comparison_force}
 \end{figure*}

Using an implementation of the numerical scheme described in~\ref{MDR contact model with a bulk elastic response numerical implementation}, comparison of the MDR contact model with a bulk response to the FEM simulations in Section~\ref{Finite element simulations showcasing a bulk elastic response} is carried out. Consistent with the FEM investigation, four types of contacts are inspected: triaxial compaction, die compaction primary contact, die compaction secondary contact, and uniaxial compression. The force-displacement curves are shown in Fig.~\ref{bulk_fem_mdrcm_comparison_force}(a)-(d). 

In the cases of triaxial and die compaction the rapid stiffening in the bulk elastic regime is properly captured by the MDR contact model with the superimposed bulk elastic response. The exact displacement that triggers the bulk response differs with $\delta_o/R_o = 0.21$ and $\delta_o/R_o = 0.49$ corresponding to triaxial and die compaction, respectively. In both cases, the same bulk response indicator value of $\psi_b = 0.08$ was used showing that it is a robust criterion not specific to one loading. As expected uniaxial compression does not show any bulk response due to lack of confinement and the MDR contact model with a bulk response predicts identical behavior as it did in Part I.

Unloading is captured nicely by the MDR contact model with a bulk response, with both the non-linearity in the curves being present as well as an accurate prediction of the unrecovered or plastic displacement. In triaxial compaction it can be seen that the accumulation of plastic displacement ceases once the bulk elastic regime begins; this is due to the fact that all displacements are equal to the mean surface displacement (i.e. the bulk displacement), thus the displacement deviator is zero. In the case of the die compaction secondary contact the force remains constant during unloading after exiting the bulk elastic response; this is because these contacts are formed by plastic deformation inducing radial expansion of the particle and not relative displacement of the contact point relative to the particle center. This plastic deformation is not reversible so the contact persists even after the primary contact is unloaded.

Most remarkably, the contact model is able to detect and accurately evolve the force at the secondary contacts. As just stated, these contacts are formed by the multi-neighbor dependent  effects of the primary contact and not from any relative displacement between the rigid flat and particle. Nothing about the location or response of these contacts is pre-programmed in the contact model. Furthermore, they correctly exhibit the bulk elastic response triggering at the same time as the primary contacts. This is a major advantage of the contact model since formation of contacts caused by multi-neighbor dependent  effects is often outright ignored in simulation of elastic-plastic particles due to inability of contact models to detect them.    

\subsection{Contact area-displacement comparison}

\begin{figure*} [!htb]
 	\centering
 	\includegraphics[width=\textwidth, trim = 1cm  2.5cm 13cm 1cm]{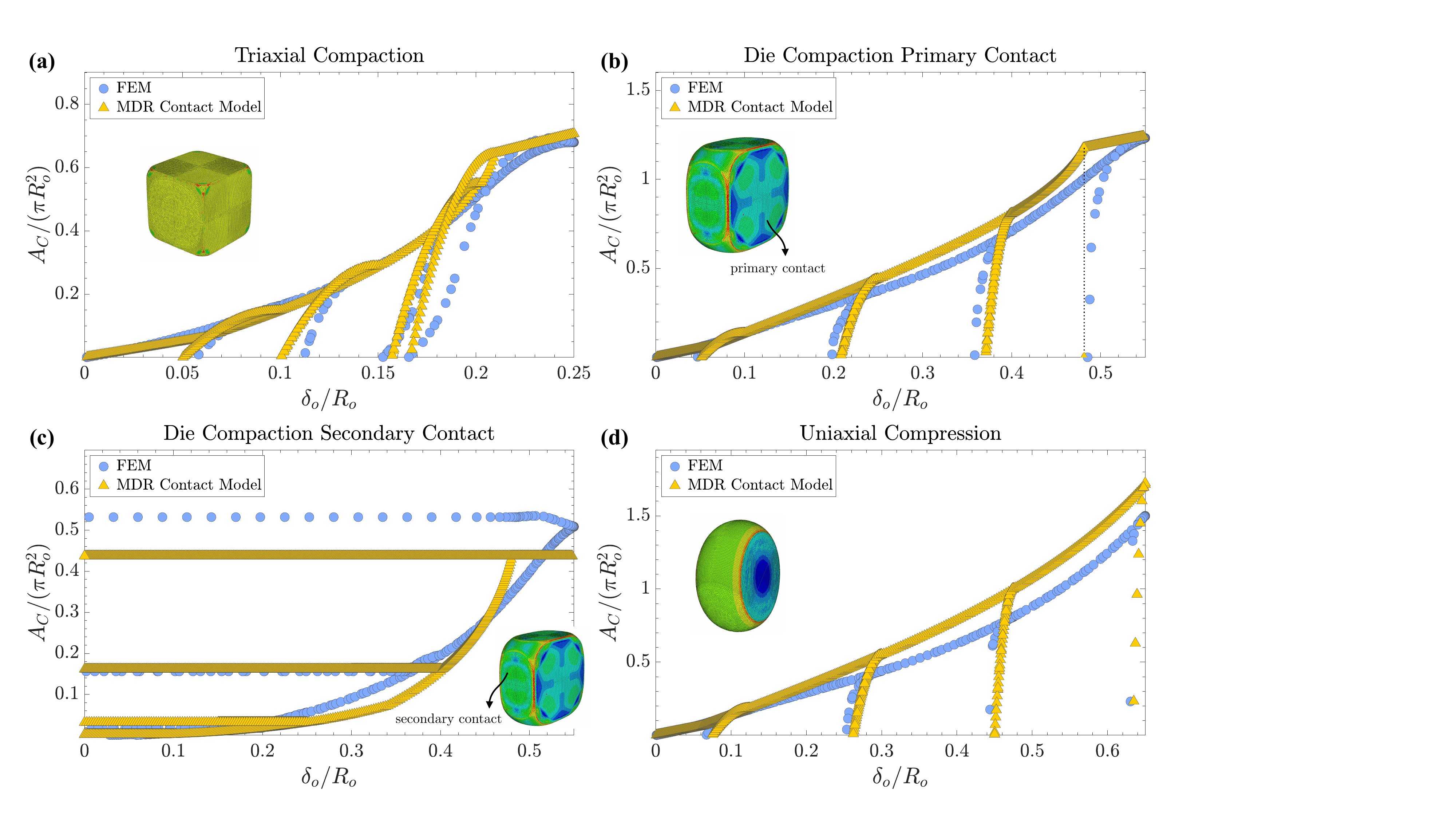}
 	\caption{Contact area-displacement curves for the four types of contacts studied: (a) triaxial compaction, (b) die compaction primary contact, (c) die compaction secondary contact, and (d) uniaxial compression. The dotted line in (b) is added to indicate the force dropping to zero.}
 	\label{bulk_fem_mdrcm_comparison_area}
 \end{figure*}

The area plotted in Fig.~\ref{bulk_fem_mdrcm_comparison_area}(a)-(d) is a composite of different formulations. Prior to the bulk response, the plotted area is identical to the formulation used in Part I: $A_C = \pi \delta_\textrm{MDR} R$ for the elastic regime, $A_C = \pi(2\delta_\textrm{MDR} R - \delta^2_\textrm{MDR}) + c_A$ for the fully plastic regime during forward loading, and the contact area predicted from the remaining springs in contact with the plane elliptical indenter in the transformed MDR space during unloading or reloading in the fully plastic regime. If the bulk response is triggered, the aforementioned areas are overridden and the area predicted by the bulk response (\ref{A_{C,i}}) is plotted instead. Although identical in form to the area used in forward loading for the fully plastic regime, (\ref{A_{C,i}}) uses the apparent overlap $\delta$ in place of $\delta_\textrm{MDR}$. 

Generally good agreement regarding the contact area is seen between the FEM simulations and the MDR contact model with a bulk response. For the case of triaxial compaction, Fig.~\ref{bulk_fem_mdrcm_comparison_area}(a), the area continues to change once the bulk response initiates. This is because $\delta$ at each contact grows with $\delta_o$, hence the geometric area predicted by (\ref{A_{C,i}}) continues to grow\footnote{Note that all area growth in the bulk elastic regime is due to change in $\delta_o$. Radius change, the other influencing factor, is not permitted while the bulk elastic regime is triggered as discussed in Section~\ref{Stopping radius growth in the bulk elastic regime}}. In the case of die compaction primary contact we see similar growth after the bulk response. Upon unloading the die compaction primary contact after the bulk response has been triggered, it is observed that the area drops abruptly from a finite value to zero. The reason for this is that the MDR and bulk components of displacement simultaneously are unloading and, in this instance, the area predicted by the remaining contacting springs in the transformed MDR space reduces to zero before (\ref{A_{C,i}}). Because (\ref{A_{C,i}}) is only valid for a nonzero bulk displacement, once the bulk displacement becomes zero the area will automatically also be predicted as zero, hence the jump in area. In the case of the die compaction secondary contact, the area growth ceases after the bulk elastic response is triggered. This is because the radius growth, which is halted during the bulk elastic response, is the only factor driving evolution of the area. Prior to the bulk response, there is a reasonably accurate evolution of the contact area at the die compaction secondary contact showing that the radius growth scheme can accurately detect and evolve contacts formed by multi-neighbor effects. The case of uniaxial compression is again consistent with the predictions from Part I.

\subsection{Average contact pressure-displacement comparison}

\begin{figure*} [!htb]
 	\centering
 	\includegraphics[width=\textwidth, trim = 1cm  2.5cm 13cm 1cm]{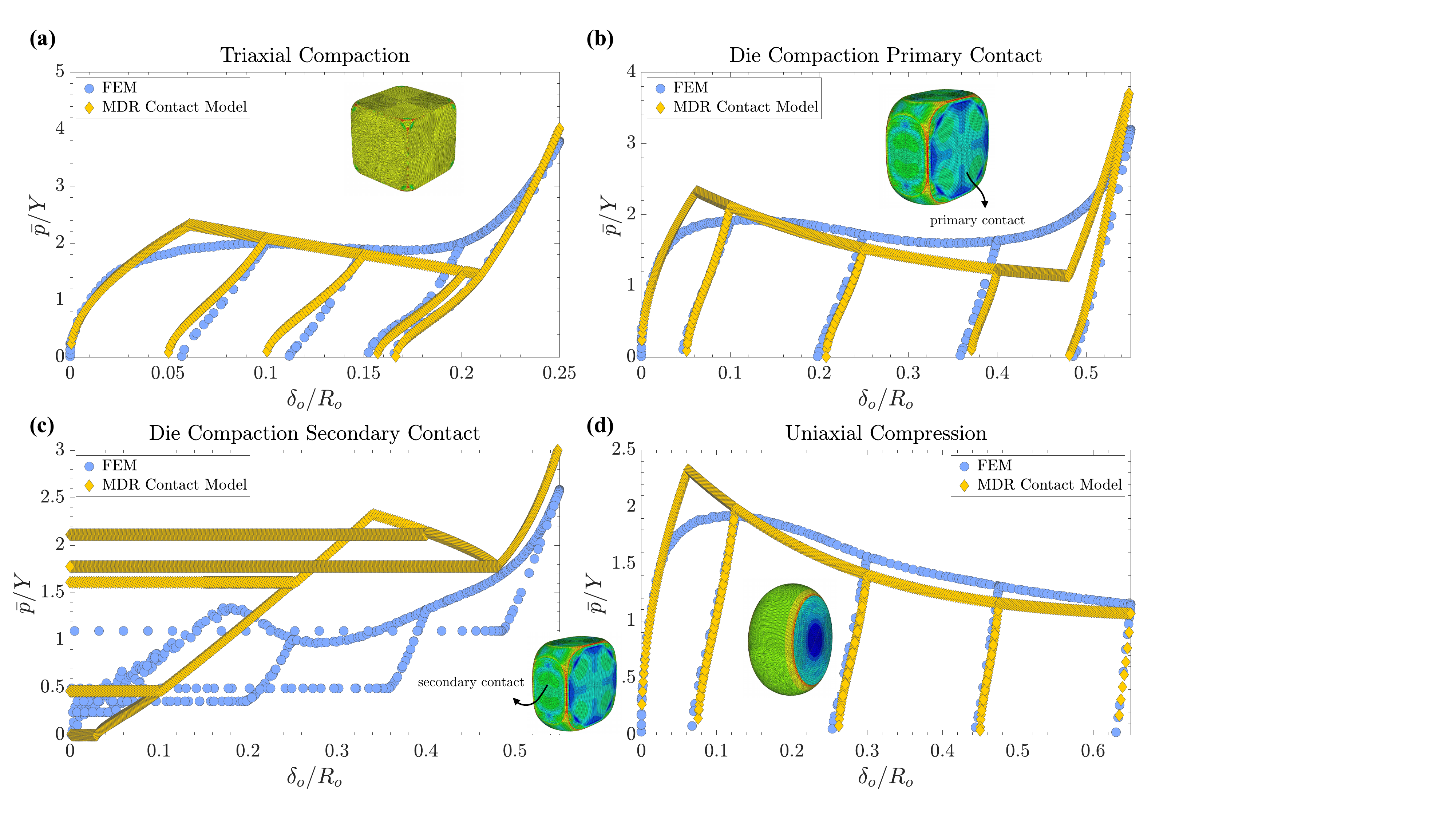}
 	\caption{Average contact pressure-displacement curves for the four types of contacts studied: (a) triaxial compaction, (b) die compaction primary contact, (c) die compaction secondary contact, and (d) uniaxial compression.}
 	\label{bulk_fem_mdrcm_comparison_pbar}
 \end{figure*}

Comparison of the average contact pressure-displacement curves between the MDR contact model with a bulk response and FEM is shown in Fig.~\ref{bulk_fem_mdrcm_comparison_pbar}(a)-(d). Generally good agreement between the two is noted. Initially we see that triaxial compaction, die compaction primary contact, and uniaxial compression all share the same evolution with an elastic (or Hertzian) average contact stress profile that then links up with the hardening curve $p_Y$. Deviation occurs once the bulk response turns on causing a sharp uptick in the average pressure curves in the case of triaxial and primary die compaction. The evolution of the die compaction secondary contact is much the same following an elastic response initially until it joins the hardening curve for a brief period, and then showing a bulk elastic response. Note that some pressure remains even after unloading due to the contact being caused by permanent plastic deformation.

\subsection{Volume-displacement comparison}

\begin{figure*} [!htb]
 	\centering
 	\includegraphics[width=\textwidth, trim = 1cm  2.5cm 13cm 1cm]{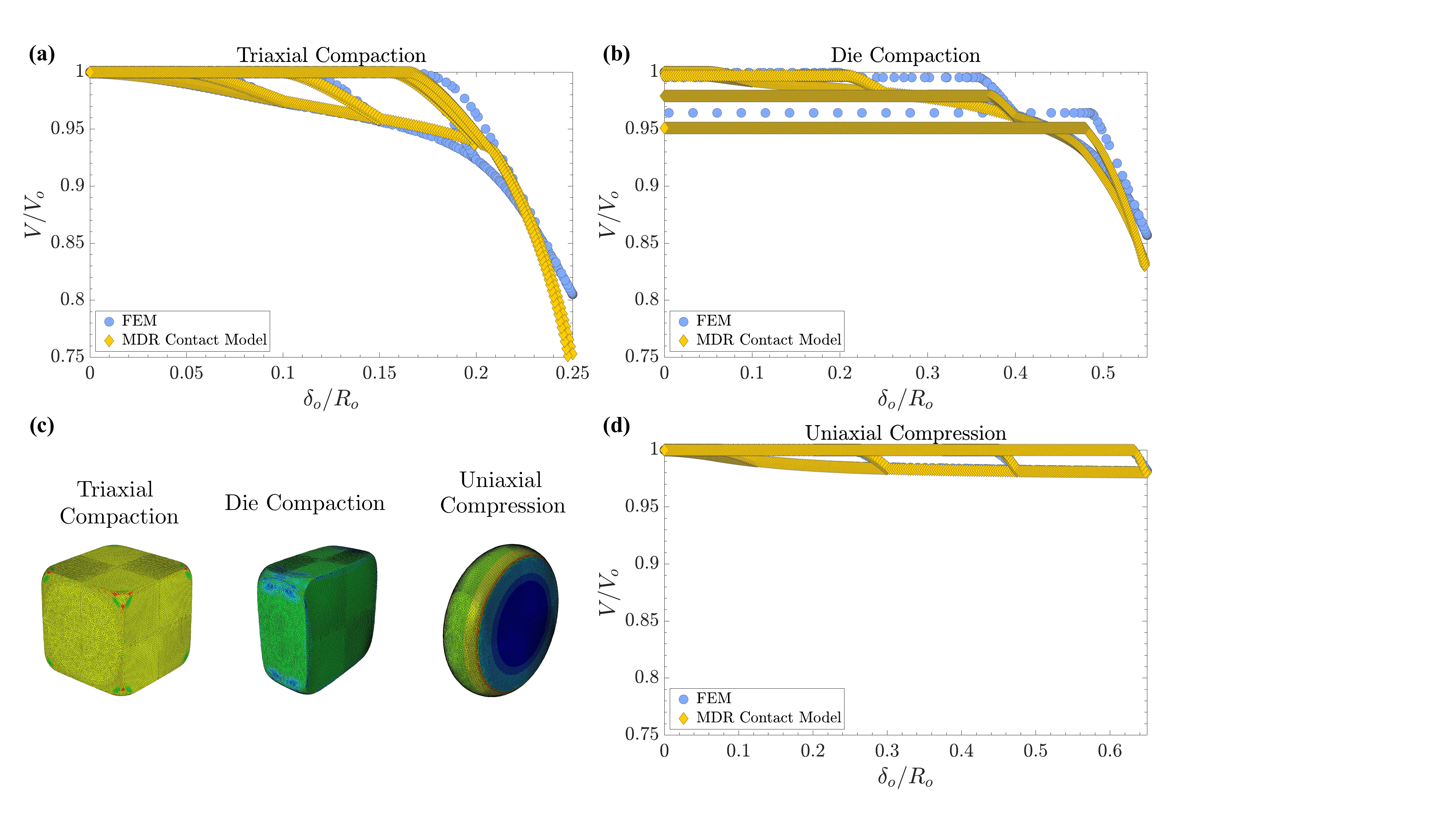}
 	\caption{Particle volume-displacement curves for the three types of loadings studied: (a) triaxial compaction, (b) die compaction, and (d) uniaxial compression. (c) Deformed particle shapes.}
 	\label{bulk_fem_mdrcm_comparison_vol}
 \end{figure*}

Comparison of the particle volume-displacement curves between the MDR contact model with a bulk response and FEM is shown in Fig.~\ref{bulk_fem_mdrcm_comparison_vol}(a)-(d). The particle volume $V$ is calculated using (\ref{elastic volume}) and is normalized by the initial volume $V_o = 4/3 \pi R_o^3$. Nice agreement is seen for the three different loading types and the vastly different volume changes are able to be correctly distinguished. For both triaxial and die compaction the rapid volume loss during the bulk elastic regime is correctly captured, which is directly linked to the increased force caused by the bulk elastic response. In the cases of triaxial compaction and uniaxial compression, unloading returns the particle to its initial volume, this is reflective of the fact that all volume change is elastic due to the incompressibility of the plastic deformation. For the case of die compaction, upon unloading the volume does not return to its initial value, this effect is caused by the forces at the secondary contacts remaining nonzero after the primary contact is unloaded. The MDR contact model with a bulk response correctly predicts this residual volume change after the primary contact is unloaded. The case of uniaxial compression is once again consistent with the results from Part I and shows no bulk elastic response.

\section{Conclusions and perspectives}

In the second part of this two part series, we have proposed a treatment to capture the force in the bulk elastic regime between contacting particles. This treatment stands independent of the MDR contact model presented in Part I, and can therefore be superimposed on any suitable existing contact model. The treatment is derived from linear elasticity and relies only on knowledge of one material property---the bulk modulus---and three calculable kinematic quantities---the mean surface displacement, contact areas, and particle volume. 

To start, the MDR contact model from Part I---lacking a bulk response---was succinctly summarized. A new suite of finite element simulations were conducted considering contact of an elastic-perfectly plastic particle and rigid flats  for three types of loadings: triaxial compaction, die compaction, and uniaxial compression. The force, contact area, average contact pressure, and particle volume behaviors as a function of displacement were analyzed. It was shown that triaxial and die compaction both lead to bulk elastic responses, while uniaxial compression does not. Die compaction was also shown to give rise to two contact types: the primary contact along the axis of compaction and four identical secondary contacts caused by multi-neighbor dependent  effects (i.e. caused by radial expansion induced by the primary contact).

In developing the bulk elastic treatment, a new simple criterion for triggering the bulk elastic response was proposed based on comparing the ratio of free surface area to total surface area. This idea led to a computationally efficient criterion that is related to the true physical mechanism that gives rise to the bulk elastic response, vanishing of pore space. A kinematic decomposition of the displacement into its bulk component and (in this case) MDR component was stated. The force related to the bulk elastic response was arrived at through relating the mean surface displacements to mean stress through some geometric properties and the bulk modulus of the material. 

Validation of the newly proposed bulk elastic response was done by superimposing it with the MDR contact model from Part I. Overall, this comparison showed that the contact model is able to recover the force, contact area, average contact pressure, and particle volume as a function of displacement with good accuracy under a variety of loadings. Importantly, the model is able to detect and evolve secondary contacts without additional effort. 

With the combination of the MDR contact model from Part I and the bulk elastic treatment here, a comprehensive contact model has been developed for spherical elastic-perfectly plastic particles. It is able to span all three regimes (elastic, fully-plastic, and bulk elastic), can recover unloading accurately at all stages without any additional or special treatment, has the capability of to model adhesive contact under proper conditions, and can detect and evolve contacts caused by multi-neighbor dependent effects. 

Worthwhile future research directions that would allow for greater accuracy in modeling more complex loadings include: adding a bulk shear response related to the shear modulus (once tangential contact modeling is included), reformulating the differential radius update scheme for high confinement, and implementing the contact model into an open-source DEM code for many interacting bodies. The last of these is the chief task needed to allow the contact model to be used for industrially relevant engineering problems and is under current development.  

\section*{Declaration of competing interest}

The authors declare that they have no known competing financial interests or personal relationships that could have appeared to influence the work reported in this paper.

\section*{Acknowledgements}

The authors acknowledge the support of the International Fine Particle Research Institute (IFPRI) (Grant ARR-109-01). The authors would like to thank the members of IFPRI for useful discussions related to the development of the contact model.

 \bibliographystyle{elsarticle-num} 
 \bibliography{bib}






\appendix
\clearpage
\section{MDR contact model with a bulk elastic response numerical implementation} \label{MDR contact model with a bulk elastic response numerical implementation}

\subsection*{Main function} 

A sketch of the numerical routine to calculate the forces at $N$ active rigid flat contacts on a single particle with consideration for a bulk elastic response is outlined below. Parameters that are underlined indicate an array of quantities for each active contact on a particle, for example, $\underline{\delta}_o$ is an array containing the displacement of each active contact on a particle. To index a specific entry of the array the subscript $i$ will be used, for example $\delta_{o,i}$ is the $i$th component of the array $\underline{\delta}_o$.

The known data is the initial radius $R_o$, composite plane strain modulus $E^*_c$, Poisson's ratio $\nu$, bulk modulus $\kappa$, yield stress $Y$, effective surface energy $\Delta\gamma$, and a value for $\psi_b$. This data is assumed to be globally available to all functions and therefore will not be explicitly listed as function inputs. The inputs at the beginning of the step are the current apparent overlap at each contact $\underline{\delta}^{n}$, current maximum MDR apparent overlap at each contact $\underline{\delta}^n_\textrm{max,MDR}$, current MDR apparent overlap at each contact $\underline{\delta}^n_\textrm{MDR}$, current bulk displacement at each contact $\underline{\delta}^n_\textrm{o,Bulk}$, current apparent radius $R^n$, current volume $V^n$, change in apparent overlap at each contact $\Delta \underline{\delta}$, and the change in displacement at each contact $\Delta \underline{\delta}_o$. Although not directly listed as inputs, it is assumed that the contact area intercept $\underline{c}_A$ and yield flag $\underline{Y}_\textrm{flag}$ (i.e. 0 = no yielding yet and 1 = yielding) at each contact are known as they are naturally calculated within calcMDRforce(...). Using the data and inputs, the numerical algorithm is constructed to output the updated apparent overlap at each contact $\underline{\delta}^{n+1}$, updated maximum MDR apparent overlap at each contact $\underline{\delta}^{n+1}_\textrm{max,MDR}$, updated MDR apparent overlap at each contact $\underline{\delta}^{n+1}_\textrm{MDR}$, updated bulk displacement at each contact $\underline{\delta}^{n+1}_\textrm{o,Bulk}$, updated apparent radius $R^{n+1}$, updated current volume $V^{n+1}$, and force at each contact $\underline{F}^{n+1}$. 

A few quantities that are taken as inputs into calcBulkItems(...) and findUpdatedRadiusAndVol(...) are not given as inputs to the main function including: the vector-valued change in displacement at each contact $\Delta \underline{\bm{\delta}}_o$, updated force vector at each contact $\underline{\bm{f}}^{n+1}$, and updated branch vector at each contact $\underline{\bm{b}}^{n+1}$. The reason for this is that these quantities can be easily determined provided the contact normal vector at each contact $\underline{\bm{n}}$ is known---as is the case for DEM simulations. Writing out their calculation explicitly would distract from the more important aspects of the numerical implementation.  

\begin{tcolorbox}[
    boxrule=3pt,        
    rounded corners,      
    width=\textwidth,    
    title=\textbf{Main function}
]

\noindent \textbf{Data}: $R_o$, $E^*_c$, $\nu$, $\kappa$, $Y$, $\Delta\gamma$, $\psi_b$ 

\noindent \textbf{Input}: $\underline{\delta}^{n}$, $\underline{\delta}^n_\textrm{max,MDR}$, $\underline{\delta}^n_\textrm{MDR}$, $\underline{\delta}^n_\textrm{o,Bulk}$, $R^n$, $V^n$, $\Delta \underline{\delta}$, $\Delta \underline{\delta}_o$

\noindent \textbf{Output}: $\underline{\delta}^{n+1}$, $\underline{\delta}^{n+1}_\textrm{max,MDR}$, $\underline{\delta}^{n+1}_\textrm{MDR}$, $\underline{\delta}^{n+1}_\textrm{o,Bulk}$, $R^{n+1}$, $V^{n+1}$, $\underline{F}^{n+1}$

\noindent \textbf{begin} 

\noindent \qquad $\underline{\delta}^{n+1} = \underline{\delta}^{n} + \Delta\underline{\delta}$

\noindent \qquad $N = \textrm{length}(\underline{\delta}^{n})$

\noindent \qquad $[\Delta \bar{\delta}_o,\underline{A}^{n+1}_C, A_\textrm{con},A_\textrm{tot},A_\textrm{free},V_\textrm{geo}] = \textrm{calcBulkItems}(R^n,N,\underline{\delta}^{n+1},\underline{\delta}^n_\textrm{max,MDR},\underline{\delta}^n_\textrm{o,Bulk},\Delta \underline{\bm{\delta}}_o, ...$

\noindent \qquad \qquad \qquad \qquad \qquad \qquad \qquad \qquad \qquad \qquad \qquad \qquad \qquad \qquad \qquad \qquad \qquad \quad $\underline{Y}_\textrm{flag},\underline{c}_A,\underline{\bm{n}})$

\noindent \qquad \textbf{for} $i = 1 : N$

        \noindent \qquad \qquad \textbf{if} $\psi_b < A_\textrm{free}/A_\textrm{tot}$
    
            \noindent \qquad \qquad \qquad $\Delta \delta_\textrm{MDR} = \Delta \delta_i$ 
    
            \noindent \qquad \qquad \qquad $\Delta \delta_{o,\textrm{Bulk}} = 0$ 
    
        \noindent \qquad \qquad \textbf{else}
    
            \noindent \qquad \qquad \qquad $\Delta \delta_\textrm{MDR} = \Delta \delta_i - \Delta \bar{\delta}_o$ 
    
            \noindent \qquad \qquad \qquad $\Delta \delta_{o,\textrm{Bulk}} = \Delta \bar{\delta}_o$ 
    
        \noindent \qquad \qquad \textbf{end} 
    
        \noindent \qquad \qquad $\delta^{n+1}_\textrm{MDR,i} = \delta^{n}_\textrm{MDR,i} + \Delta \delta_\textrm{MDR}$ 
    
        \noindent \qquad \qquad $\delta^{n+1}_\textrm{o,Bulk,i} = \textrm{max} [  \delta^{n}_\textrm{o,Bulk,i} + \Delta \delta_\textrm{o,Bulk},\;0]$ 
    
        \noindent \qquad \qquad \textbf{if} $\delta^{n+1}_\textrm{MDR,i} > \delta^n_\textrm{max,MDR,i}$ 
    
            \noindent \qquad \qquad \qquad $\delta^{n+1}_\textrm{max,MDR,i} = \delta^{n+1}_\textrm{MDR,i}$ 
    
        \noindent \qquad \qquad \textbf{else}
    
            \noindent \qquad \qquad \qquad $\delta^{n+1}_\textrm{max,MDR,i} = \delta^n_\textrm{max,MDR,i}$
    
        \noindent \qquad \qquad \textbf{end}
    
        \noindent \qquad \qquad $F_\textrm{MDR} = \textrm{calcMDRforce}(\delta^{n+1}_\textrm{MDR,i},\delta^{n+1}_\textrm{max,MDR,i},R^{n},c_{A,i},Y_\textrm{flag,i})$
    
        \noindent \qquad \qquad $F_\textrm{Bulk} = \frac{A_\textrm{con}}{V_\textrm{geo}} \delta^{n+1}_\textrm{o,Bulk,i} \kappa A^{n+1}_{C,i}$   
        
        \noindent \qquad \qquad $F^{n+1}_i = F_\textrm{MDR} + F_\textrm{Bulk}$

    \noindent \qquad \textbf{end}

    \noindent \qquad $[R^{n+1},V^{n+1}] = \textrm{findUpdatedRadiusAndVol}(R^n,N,\underline{\delta}^{n+1}_o,\Delta \underline{\delta}_o,\underline{\bm{f}}^{n+1},\underline{\bm{b}}^{n+1},V^n)$

\noindent \textbf{end}
\end{tcolorbox} 

\subsection*{Functions to calculate the mean displacement and updated radius}

To complete the numerical implementation calcBulkItems(...) and findUpdatedRadiusAndVol(...) require definition. Both the case of finding the mean displacement and updated contact radius are multi-neighbor dependent  processes, meaning that they source information from all contacts on a given particle. It is important to highlight the fact that both functions exist outside the \textit{for} loop that runs over each active contact, this is because all outputs of the functions are particle properties, excluding $\underline{A}^{n+1}_C$. We begin with calcBulkItems(...). \\

\begin{tcolorbox}[
    boxrule=1pt,        
    rounded corners,      
    width=\textwidth    
]

\noindent calcBulkItems($R^n$,$N$,$\underline{\delta}^{n+1}$,$\underline{\delta}^n_\textrm{max,MDR}$,$\underline{\delta}^n_\textrm{o,Bulk}$,$\Delta \underline{\bm{\delta}}_o$,$\underline{Y}_\textrm{flag}$,$\underline{c}_A$,$\underline{\bm{n}}$)

\tcblower

    \noindent \qquad \textbf{for} $i = 1 : N$

        \noindent \qquad \qquad \textbf{if} $Y_{\textrm{flag},i} = 0$

            \noindent \qquad \qquad \qquad $A^{n+1}_{C,i} = \pi R^n \delta^{n+1}_i$

        \noindent \qquad \qquad \textbf{else} 
    
            \noindent \qquad \qquad \qquad $A^{n+1}_{C,i} = \pi (2 R^n \delta^{n+1}_i - (\delta^{n+1}_i)^2)  + c_{A,i}$

        \noindent \qquad \qquad \textbf{end}
        
    \noindent \qquad \textbf{end}

    \noindent \qquad $A_\textrm{con} = \sum_{i=1}^{N} A^{n+1}_{C,i}$ 

    \noindent \qquad $\Delta \bar{\delta}_o = \frac{1}{A_\textrm{con}} \sum_{i=1}^{N} (A^{n+1}_{C,i} \bm{n}_i \cdot \Delta \bm{\delta}_{o,i})$

    \noindent \qquad $A_\textrm{tot} = 4\pi (R^n)^2 - 2\pi \sum_{i=1}^{N} (\delta^n_\textrm{max,MDR,i} + \delta^n_{o\textrm{,Bulk,i}})R^n + A_\textrm{con}$

    \noindent \qquad $A_\textrm{free} = A_\textrm{tot} - A_\textrm{con}$

    \noindent \qquad $V_\textrm{geo} = \frac{4}{3}\pi (R^n)^3 - \sum_{i=1}^{N} \left( \frac{\pi}{3} (\delta^{n+1}_i)^2[3R^n - \delta^{n+1}_i] \right)$

\noindent \textbf{end}

\end{tcolorbox}

\noindent To find the updated particle radius an explicit numerical scheme is used that ensures all volume change during a particular step is due to elastic deformation. 

\begin{tcolorbox}[
    boxrule=1pt,        
    rounded corners,      
    width=\textwidth    
]

\noindent findUpdatedRadiusAndVol($R^n$,$N$,$\underline{\delta}^{n+1}_o$,$\Delta \underline{\delta}_o$,$\underline{\bm{f}}^{n+1}$,$\underline{\bm{b}}^{n+1}$,$V^n$)

\tcblower

    \noindent \qquad $V_o = 4/3 \pi R^3_o$ 
    
    \noindent \qquad $V^{n+1} = V_o \left[ 1 + \frac{1}{3\kappa V^{n}}\textrm{tr} \left( \sum^N_{i = 1} \bm{f}^{n+1}_i \otimes \bm{b}^{n+1}_i \right) \right]$  

    \noindent \qquad $\Delta V^e = -(V^{n+1} - V^n)$

    \noindent \qquad $\Delta R = \textrm{max} \left[ \frac{\Delta V^e - \sum_{i=1}^{N} \pi\Delta \delta_{o,i}(2\delta^{n+1}_{o,i} R_o - (\delta^{n+1}_{o,i})^2 + (R^n)^2 - R^2_o) }{2 \pi R^n \sum_{i=1}^{N}(\delta^{n+1}_{o,i} + R^n - R_o) - 4 \pi (R^n)^2}, \; 0 \right]$

    \noindent \qquad \textbf{if} $\psi_b < A_\textrm{free}/A_\textrm{tot}$
    
        \noindent \qquad \qquad $R^{n+1} = R^n + \Delta R$  
        
    \noindent \qquad \textbf{else}
    
        \noindent \qquad \qquad $R^{n+1} = R^n$
        
    \noindent \qquad \textbf{end}

\noindent \textbf{end} 

\end{tcolorbox}

\subsection*{MDR force calculation function}

The code to calculate the MDR contribution to the total force $F_\textrm{MDR}$ is identical to that given in Part I. The apparent overlaps in this case are from the MDR portion of displacement. 

\begin{tcolorbox}[
    boxrule=1pt,        
    rounded corners,      
    width=\textwidth    
]

\noindent calcMDRforce($\delta^{n+1}_\textrm{MDR,i}$,$\delta^{n+1}_\textrm{max,MDR,i}$,$R^{n}$,$c_{A,i}$,$Y_\textrm{flag,i}$)

\tcblower

    \noindent \qquad $\delta = \delta^{n+1}_\textrm{MDR,i}$

    \noindent \qquad $\delta_\textrm{max} = \delta^{n+1}_\textrm{max,MDR,i}$

    \noindent \qquad $R = R^{n}$

    \noindent \qquad $p_Y = Y\left( 1.75\exp{(-4.4\delta_\textrm{max}/R)+1} \right)$

    \noindent \qquad $\bar{p}_H = \frac{4E^*_c}{3 \pi \sqrt{R}}\sqrt{\delta}$

    \noindent \qquad \textbf{if} $Y_\textrm{flag,i} = 0$ and $\bar{p}_H > p_Y$

        \noindent \qquad \qquad $\delta = \delta_Y$
        
        \noindent \qquad \qquad $c_{A,i} = \pi(\delta_Y^2 - \delta_Y R)$
        
        \noindent \qquad \qquad $Y_\textrm{flag,i} = 1$
    
    \noindent \qquad \textbf{end}

    \noindent \qquad \textbf{if} $Y_\textrm{flag,i} = 0$ 
    
        \noindent \qquad \qquad $A = 4R$

        \noindent \qquad \qquad $B = 2R$

        \noindent \qquad \qquad $\delta^e_\textrm{1D} = \delta$
        
    \noindent \qquad \textbf{else}

        \noindent \qquad \qquad $a_\textrm{max} = \sqrt{(2\delta_\textrm{max} R - \delta^2_\textrm{max}) + c_{A,i}/\pi}$

        \noindent \qquad \qquad $A = \frac{4p_Y}{E^*_c}a_\textrm{max}$

        \noindent \qquad \qquad $B = 2a_\textrm{max}$

        \noindent \qquad \qquad $\delta^e_\textrm{1D,max} = A/2$

        \noindent \qquad \qquad $F_\textrm{max} = \frac{E^*_c AB}{4}\left[ \arccos\left( {1 - \frac{2\delta^e_{\textrm{1D,max}}}{A}}\right) - \left( 1 - \frac{2\delta^e_{\textrm{1D,max}}}{A} \right) \sqrt{\frac{4\delta^e_{\textrm{1D,max}}}{A} - \frac{4(\delta^e_{\textrm{1D,max}})^2}{A^2}}\right]$

        \noindent \qquad \qquad $z_R = R - (\delta_\textrm{max} - \delta^e_\textrm{1D,max})$

        \noindent \qquad \qquad $\delta_R = \frac{F_\textrm{max}}{\pi a_\textrm{max}^2} \left[  \frac{2a^2_\textrm{max} (\nu - 1) - z_R(2\nu - 1)( \sqrt{a^2_\textrm{max} + z^2_R} - z_R)} {2G\sqrt{a_\textrm{max}^2 + z_R^2}} \right]$

        \noindent \qquad \qquad $\delta^e_\textrm{1D} = \frac{\delta - \delta_\textrm{max} + \delta^e_\textrm{1D,max} + \delta_R}{1 + \delta_R/\delta^e_\textrm{1D,max}}$
        
    \noindent \qquad \textbf{end} 

    \noindent \qquad \textbf{if} $\textrm{adhesion} = \textrm{`on'}$ 

        \noindent \qquad \qquad $F_\textrm{MDR} = \textrm{forceAdhesive}(A,B,\delta^e_\textrm{1D},a)$ 

    \noindent \qquad \textbf{else}

            \noindent \qquad \qquad \textbf{if} $\delta^e_\textrm{1D} > 0$
            
                \noindent \qquad \qquad \qquad $F_\textrm{MDR} = \frac{E^*_c AB}{4}\left[ \arccos\left( {1 - \frac{2\delta^e_{\textrm{1D}}}{A}}\right) - \left( 1 - \frac{2\delta^e_{\textrm{1D}}}{A} \right) \sqrt{\frac{4\delta^e_{\textrm{1D}}}{A} - \frac{4(\delta^e_{\textrm{1D}})^2}{A^2}}\right]$

            \noindent \qquad \qquad \textbf{else} 

                \noindent \qquad \qquad \qquad $F_\textrm{MDR} = 0$

            \noindent \qquad \qquad \textbf{end} 

    \noindent \qquad \textbf{end}

\noindent \textbf{end} 

\end{tcolorbox}

To model adhesive contact forceAdhesion(...) requires definition. Unlike the normal contact case the contact radius $a$ must be explicitly tracked during adhesive contact; we assume it is a known input whose value gets updated within forceAdhesion(...) and passed between steps. The subscripts $\textrm{n.a.}$ and $\textrm{a.r.}$ stand for non-adhesive and adhesive retraction, respectively. A full description of the theory behind the adhesive contact is provided in Part I.  

\begin{tcolorbox}[
    boxrule=1pt,        
    rounded corners,      
    width=\textwidth    
]

\noindent forceAdhesion($A$,$B$,$\delta^e_\textrm{1D}$,$a$)

\tcblower

    \noindent \qquad $g_\textrm{1D}(a) = \frac{A}{2} - \frac{A}{B}\sqrt{\frac{B^2}{4} - a^2}$

    \noindent \qquad $w_\textrm{1D}(a) = \delta^e_\textrm{1D} - g_\textrm{1D}(a)$
    
    \noindent \qquad \textbf{if} $w_\textrm{1D}(a) = 0$

        \noindent \qquad \qquad $a = B/2$
    
        \noindent \qquad \qquad $F^{n+1} = \frac{E^*_c AB}{4}\left[ \arccos\left( {1 - \frac{2\delta^e_{\textrm{1D}}}{A}}\right) - \left( 1 - \frac{2\delta^e_{\textrm{1D}}}{A} \right) \sqrt{\frac{4\delta^e_{\textrm{1D}}}{A} - \frac{4(\delta^e_{\textrm{1D}})^2}{A^2}}\right]$ 
        
    \noindent \qquad \textbf{else}

        \noindent \qquad \qquad $\Delta l = \sqrt{\frac{2 \pi a \Delta \gamma}{E^*_c}}$

        \noindent \qquad \qquad $a_c = \textrm{Solve}\left[ \frac{dg_\textrm{1D}(a)}{da}\Bigr|_{a=a_c} = \sqrt{\frac{\pi \Delta \gamma}{2 E^*_c a_c}}, \; a_c \right]$

    \noindent \qquad \qquad \textbf{if} $w_\textrm{1D}(a) < \Delta l$

        \noindent \qquad \qquad \qquad $\delta^e_\textrm{1D,adh} = g_\textrm{1D}(a)$

        \noindent \qquad \qquad \qquad $F_\textrm{n.a.} = \frac{E^*_c AB}{4}\left[ \arccos\left( {1 - \frac{2\delta^e_{\textrm{1D,adh}}}{A}}\right) - \left( 1 - \frac{2\delta^e_{\textrm{1D,adh}}}{A} \right) \sqrt{\frac{4\delta^e_{\textrm{1D,adh}}}{A} - \frac{4(\delta^e_{\textrm{1D,adh}})^2}{A^2}}\right]$ 

        \noindent \qquad \qquad \qquad $F_\textrm{a.r.} = 2 E^*_c(\delta^e_{1D} - g_\textrm{1D}(a))a$

        \noindent \qquad \qquad \qquad $F_\textrm{MDR} = F_\textrm{n.a.} + F_\textrm{a.r.}$ 

    \noindent \qquad \qquad \textbf{elseif} $w_\textrm{1D}(a) > \Delta l$

        \noindent \qquad \qquad \qquad $a = \textrm{Solve}\left[\delta^e_\textrm{1D} + \Delta l - g_\textrm{1D}(a) = 0, \; a  \right]$

        \noindent \qquad \qquad \qquad \textbf{if} $a < a_c$

            \noindent \qquad \qquad \qquad \qquad $F_\textrm{MDR} = 0$

        \noindent \qquad \qquad \qquad \textbf{else} 

            \noindent \qquad \qquad \qquad \qquad $\delta^e_\textrm{1D,adh} = g_\textrm{1D}(a)$

            \noindent \qquad \qquad \qquad \qquad $F_\textrm{n.a.} = \frac{E^*_c AB}{4}\left[ \arccos\left( {1 - \frac{2\delta^e_{\textrm{1D,adh}}}{A}}\right) - \left( 1 - \frac{2\delta^e_{\textrm{1D,adh}}}{A} \right) \sqrt{\frac{4\delta^e_{\textrm{1D,adh}}}{A} - \frac{4(\delta^e_{\textrm{1D,adh}})^2}{A^2}}\right]$ 

            \noindent \qquad \qquad \qquad \qquad $F_\textrm{a.r.} = 2 E^*_c(\delta^e_{1D} - g_\textrm{1D}(a))a$

            \noindent \qquad \qquad \qquad \qquad $F_\textrm{MDR} = F_\textrm{n.a.} + F_\textrm{a.r.}$

    \noindent \qquad \qquad \qquad \textbf{end}

    \noindent \qquad \qquad \textbf{end} 

    \noindent \qquad \textbf{end} 

\noindent \textbf{end} 

\end{tcolorbox}

\end{document}